\documentclass{article}

\usepackage{PRIMEarxiv}
\usepackage[T1]{fontenc}    
\usepackage{hyperref}       
\usepackage{url}            
\usepackage{booktabs}       
\usepackage{amsfonts}       
\usepackage{nicefrac}       
\usepackage{microtype}      
\usepackage{lipsum}
\usepackage{fancyhdr}       
\usepackage{graphicx}       
\graphicspath{{media/}}     
\usepackage{float}
\usepackage{amsmath}
\usepackage{indentfirst}
\usepackage{marvosym}

\title{Quantum Dynamics of Machine Learning
	
	}
	
	\author{
		Peng Wang \textsuperscript{\Letter} \\
		Chengdu Institution of Computer Application\\
		Chinese Academy of Sciences \\
		Chengdu, 610213, Sichuan, China.\\
		\Letter { } \texttt{wp002005@163.com}  \\
		\And
		Maimaitiniyazi Maimaitiabudula \\
		Southwest Minzu University \\
		Chengdu, 610213, Sichuan, China.\\
		\texttt{mai1232021@163.com} \\
	}
	
	\makeatletter
	\def\thanks#1{\protected@xdef\@thanks{\@thanks
			\protect\footnotetext{#1}}}
	\makeatother
	
\begin{document}
		\maketitle
		\vspace*{-20pt}
		\begin{abstract}
		The quantum dynamic equation (QDE) of machine learning is obtained based on Schrödinger equation and potential energy equivalence relationship. Through Wick rotation, the relationship between quantum dynamics and thermodynamics is also established in this paper. This equation reformulates the iterative process of machine learning into a time-dependent partial differential equation with a clear mathematical structure, offering a theoretical framework for investigating machine learning iterations through quantum and mathematical theories. Within this framework, the fundamental iterative process, the diffusion model, and the Softmax and Sigmoid functions are examined, validating the proposed quantum dynamics equations. This approach not only presents a rigorous theoretical foundation for machine learning but also holds promise for supporting the implementation of machine learning algorithms on quantum computers.
		\end{abstract}
		\section{Introduction}
		Machine learning is a typical optimisation problem, and its learning process is an iterative optimisation process in the parameter space. It is a natural way of thinking to consider the iterative motion process of this algorithm as a kinetic process. The theory of dynamics has been developed over a long period of time and is very complete, with quantum dynamics, Newtonian dynamics, thermodynamics, electrodynamics and molecular dynamics, which theoretically describes the laws of motion by establishing a set of kinetic equations. The establishment of the dynamics theory of machine learning is expected to address the lack of theoretical models in this field, thus advancing the development of machine learning theory and applications.\\
		The theoretical modeling of optimization problems using a dynamics approach was pioneered by Metropolis in 1953, drawing from thermodynamics \cite{metropolis1953equation}. This concept was later adapted by Kirkpatrick et al. for the simulated annealing algorithm, an early attempt to apply thermodynamics to optimization \cite{kirkpatrick1983optimization}. In 1994, Finnila et al. introduced a groundbreaking method that treated the objective function as a potential within the Schrödinger equation, transforming the optimization problem into one of finding a bound-state quantum ground state wave function. This marked the first application of quantum dynamics theory to optimization. In 2013, our research began exploring the potential of a quantum dynamics framework for optimization problems. Our results demonstrated that the Schrödinger equation can effectively describe the fundamental iterative process of optimization algorithms \cite{wang2013multiscale,wang2022review,wang2023quantum,wang2024convergence}. Further evidence of the successful application of quantum dynamics in artificial intelligence includes the D-Wave quantum computer, which utilized quantum annealing to solve an intelligent optimization problem, marking a milestone in commercial quantum computing \cite{johnson2011quantum}.\\
		In 2015, Jascha Sohl-Dickstein and colleagues introduced the Diffusion Probabilistic Model, inspired by non-equilibrium thermodynamic principles\cite{sohl2015deep}. This model has since undergone rapid development and found extensive applications. The Diffusion Model exemplifies the successful application of kinetic theory to machine learning. Recently, dynamics theory has shown growing theoretical and practical value in this field. New methods based on dynamics, such as those utilizing stochastic differential equations, have been developed and consistently validated through experimental results \cite{song2020score}.\\
		The cornerstone of the kinetic theory of machine learning lies in formulating the kinetic equations that govern the learning process. Quantum dynamics, rooted in the fundamental laws of motion in the material world, offers a compelling framework for this purpose. The Schrödinger equation, the core of quantum dynamics, elegantly captures the probabilistic laws of motion through a deterministic, time-dependent partial differential equation. This alignment suggests that quantum dynamics could provide a robust framework for describing machine learning processes. In this paper, we explore the iterative process of machine learning through the lens of quantum dynamics, aiming to establish a quantum dynamics equation tailored for machine learning.
		\section{Quantum Dynamics Framework for Optimization Problems}
		The introduction of the Multi-scale Quantum Harmonic Oscillator Algorithm (MQHOA) in 2013 marked the inception of a quantum dynamics framework for optimization problems. Since then, a sustained research effort has been devoted to advancing this framework, resulting in significant theoretical and practical achievements in functional optimization \cite{xin2021multiscale, jin2021multiscale}.
		\begin{figure}[H]
			\centering
			\includegraphics[width=\linewidth, height=0.6\textheight]{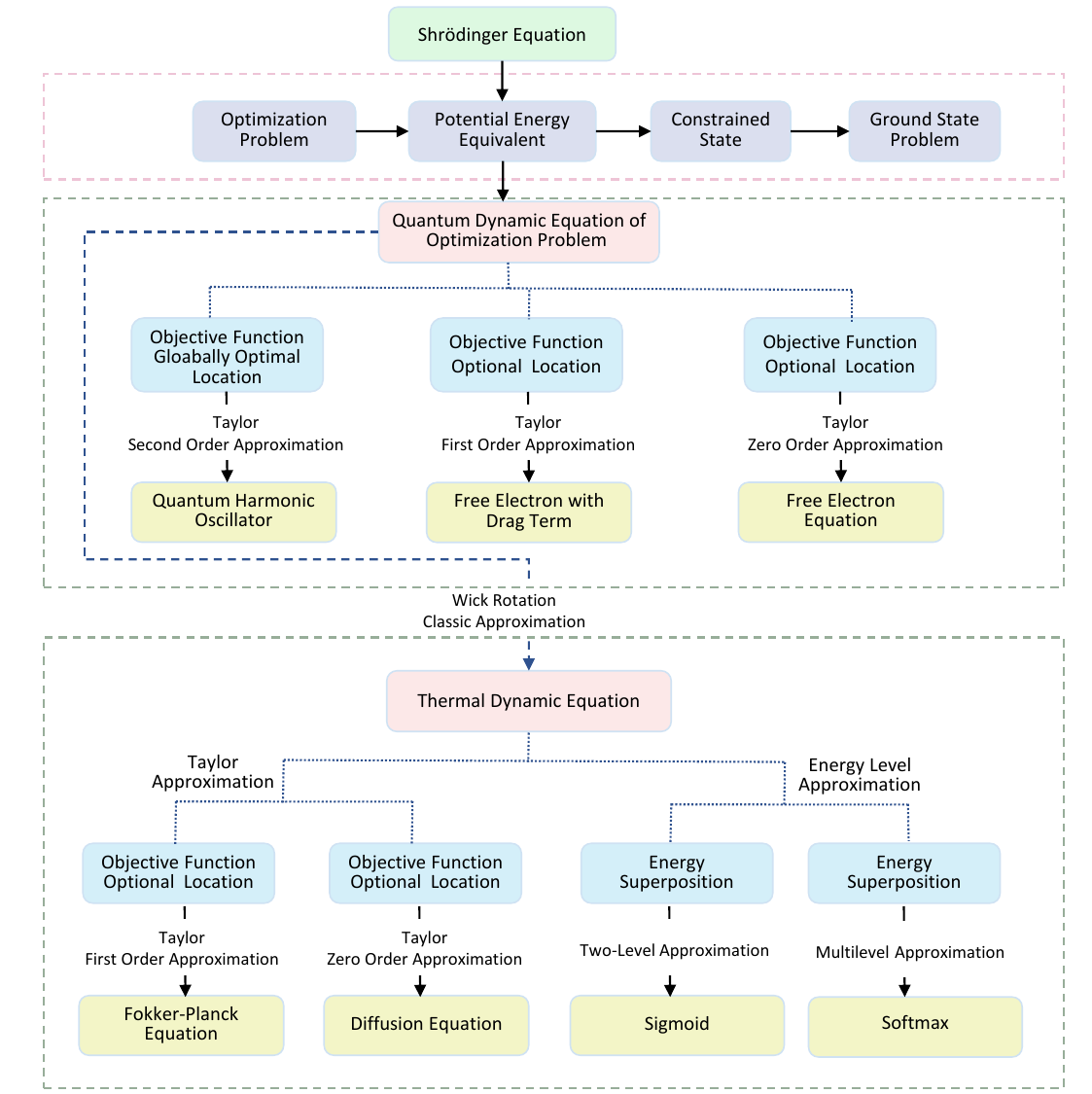}
			\caption{Quantum Dynamic Framework of Optimization Problem}
			\label{fig:1}
		\end{figure}
		The fundamental structure of the Quantum Dynamics Framework (QDF) is illustrated in Figure 1. At the core of quantum mechanics lies the Schrödinger equation. In the context of optimization problems, it is feasible to disregard the physical quantities and constants typically present in the Schrödinger equation. This simplification leads to the following form of the Schrödinger equation:
		\begin{equation}
			i\frac{\partial\psi(x,t)}{\partial t}=\Bigg[-D\frac{\partial^2}{\partial x^2}+V\big(x\big)\Bigg]\psi\big(x,t\big)
		\end{equation}
		The constant $\mathit{D}$ influences the magnitude of the system's kinetic energy, while the bound
		potential energy,$V(x)$, and the particle's wave function, $\psi(x,t)$,are also pertinent. The core concept of the Quantum Dynamics Framework (QDF) is to interpret the iterative process of an optimization problem as a quantum dynamics process, with the objective function $f(x)$ in the optimization problem considered as the potential energy term $V(x)$ in the
		Schrödinger equation. This leads to the equivalence $V(x)=f(x)$, thereby transforming the optimization problem into one of solving quantum constraints. Specifically, this approach seeks the ground state wave function of a quantum bound state, with the corresponding quantum dynamical equations for the optimization problem expressed as follows:\\
		The time-dependent Schrödinger equation for the wave function is:
		\begin{equation}
			i\frac{\partial \psi \left( x,t \right)}{\partial t} = \left[ -D\frac{{{\partial }^{2}}}{\partial {{x}^{2}}}+V\left( x \right) \right]\psi \left( x,t \right)
		\end{equation}
		This equation formulates the iterative process of the optimization problem as a partial differential equation. The time evolution of the wave function $\psi(x,t)$ represents the evolution of the probability distribution of the solution to the optimization problem.
		\section{Quantum Dynamics of Machine Learning}
		\subsection{Machine Learning Quantum Dynamics Equations}
		The process of machine learning involves optimizing parameters within a parameter space
		composed of numerous connection weights. The optimization objective is to find an
		optimal combination of these weights. Directly obtaining the objective function in this
		context is not feasible. To distinguish it from the objective function $f(x)$ in function optimization problems, we use the generalized objective function $\zeta(x)$ to formally represent the mathematical relationship of the objective function for any problem within the neural network's parameter space. This is regarded as a potential energy term in the Schrödinger equation. Thus, the potential term in the Schrödinger equation is expressed as
		\begin{equation}
			V(x) = \zeta(x)
		\end{equation}
		The Schrödinger equation is thereby transformed into the following form:
		\begin{equation}
			i\frac{\partial \psi(x,t)}{\partial t} = \left[ -D\frac{\partial^2}{\partial x^2} + \zeta(x) \right] \psi(x,t)
		\end{equation}
		This equation represents the quantum dynamics equation for machine learning, which can
		be interpreted as a system of quantum bound states defined by a potential energy
		function, $\zeta(x).$ According to Born's probabilistic interpretation of the wave function, the
		modulus squared of the wave function at a given point in time, $|\psi(x,t)|^2$, may be
		regarded as a probability distribution representing the likelihood of finding the solution at
		that point in time. The process of machine learning can thus be represented by the time
		evolution of the wave function, $\psi(x,t).$
		The Hamiltonian operator of the system is given by:
		\begin{equation}
			\hat{H} = -D\frac{\partial^2}{\partial x^2} + \zeta(x)
		\end{equation}
		The parameter $D$ determines the magnitude of the system's kinetic energy. The system can
		be annealed by gradually decreasing the value of $D$.Accordingly, the quantum dynamics
		equation of machine learning predicts that the iterative process may involve two majon iterative loops. The first loop corresponds to the quantum annealing process, where the value of $D$ is decreased. This is analogous to gradually reducing the sampling step size of
		$\zeta(x)$ in the parameter space for the generalized objective function. The second loop involves
		the time evolution of the system towards the ground state under a fixed kinetic energy condition.\\
		When the value of $D$ is significant, the system's kinetic energy is high, and the zero-point energy $(E_0)$ in the ground state is also substantial. To obtain an accurate global optimal solution, $D$ must be gradually reduced to achieve a lower zero-point energy.\\
		The significance of establishing quantum dynamics equations for machine learning is
		twofold. Firstly, it reinterprets the optimal learning process of machine learning on
		parameters in physical terms, transforming it into a problem of solving the ground state
		wave function of a quantum bound state with the generalized objective function $\zeta(x)$ as the
		constraint potential. Secondly, it mathematically converts the iterative process of machine
		learning into rigorous time-dependent differential equations. Given the century-long
		development of quantum physics into a comprehensive and intricate system of physical and
		mathematical theories, it can offer valuable insights that enhance the study of machine
		learning.
		\subsection{Taylor Approximation of the Generalized Objective Function}
		Given the complexity of solving the quantum dynamics equations analytically and the challenge of obtaining the analytical form of the generalized objective function $\zeta(x)$ in the machine learning parameter space, the quantum dynamics equations are approximated using a Taylor expansion for the generalized objective function.
		
		1. Zeroth-Order Taylor Approximation of the Generalized Objective Function
		
		In parameter space, the Taylor zeroth-order approximation of the generalized objective function, denoted by $\zeta(x)$, at the current position $x_0$ is $\zeta(x)=0.$ This implies that the generalized objective function is not defined at the current position. Consequently, the quantum dynamics equation for machine learning reduces to the free electron equation under the zeroth-order Taylor approximation of the generalized objective function $\zeta(x).$
		\begin{equation}
			i\frac{\partial \psi(x,t)}{\partial t} = -D\frac{\partial^2 \psi(x,t)}{\partial x^2}
		\end{equation}
		The motion of free electrons is unconstrained by any potential energy, hence there is no
		potential energy term, $\zeta(x)$, in the free electron equation. This equation includes an
		additional imaginary unit, $i$, which accounts for fluctuations beyond those described by the classical diffusion equation. It characterizes the quantum dynamics of an algorithm when the information about the generalized objective function, $\zeta(x)$, is entirely unknown in the machine learning process.
		\begin{figure}[H]
			\centering
			\includegraphics[width=\linewidth]{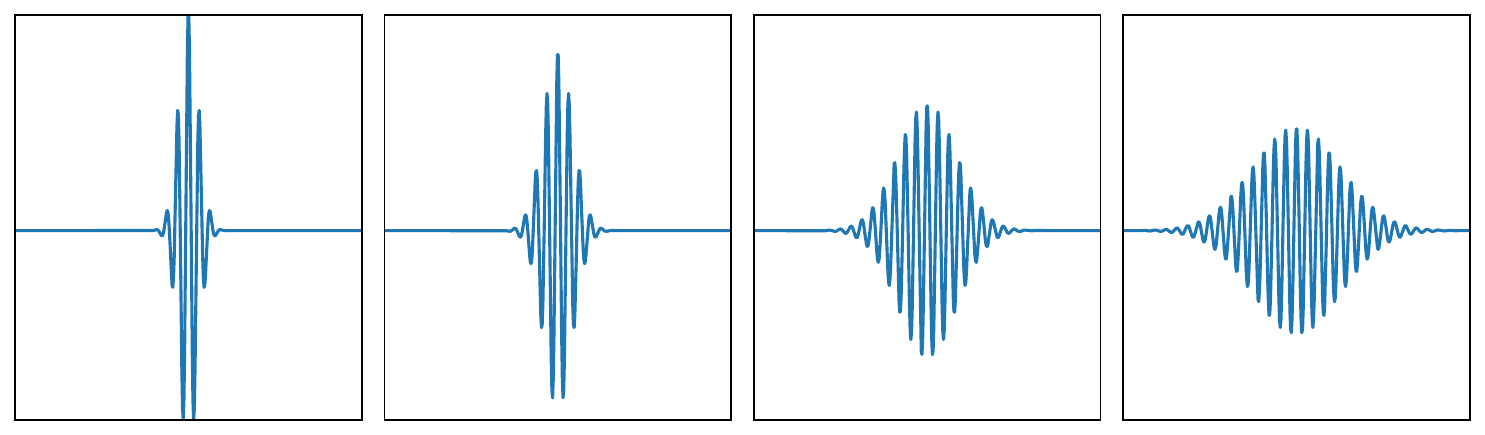}
			\caption{Wave Packet Dispersion Process}
			\label{fig:2}
		\end{figure}
		The theory of free electron motion in quantum mechanics suggests that the quantum dynamics process of machine learning, when approximated to the zeroth-order Taylor series, manifests as a wave packet dispersion (WPD) process. This dispersion results from the wave function of a free particle, which is a wave packet composed of states with various frequency components. Over time, these components gradually disperse (see Fig. 2). This process, akin to diffusion, is a more intricate wave packet dispersion, arising from the coexistence of wave-particle duality in quantum dynamical systems.
		
		2. First-Order Taylor Approximation of the Generalized Objective Function
		
		The first-order Taylor approximation of the generalised objective function, denoted by $\zeta \left( x \right)$, in parameter space at the current position ${{x}_{0}}$ is given by $\zeta \left( x \right) \approx \frac{\partial \zeta \left( {{x}_{0}} \right)}{\partial x}$. At this point, the quantum dynamics equation for machine learning is:
		\begin{equation}
			i\frac{\partial \psi(x,t)}{\partial t} = \left( -D\frac{\partial^2}{\partial x^2} + \frac{\partial \zeta(x_0)}{\partial x} \right) \psi(x,t)
		\end{equation}
		This equation represents the gradient of the wave packet as the electron moves to the
		position $x_0$, obtaining the current constraint potential $\zeta(x)$ as $\frac{\partial \zeta \left( {{x}_{0}} \right)}{\partial x}$.This dynamic mirrors the algorithmic process of gradient descent optimization. The optimization operation involves sampling the parameter space twice to estimate the slope of the generalized objective function at the position $x_0.$
		\subsection{Classical Approximation of Machine Learning Quantum Dynamics Equations}
		The process of simulating wave packet dynamics in quantum mechanics poses significant challenges for classical computers. To address this, we propose leveraging machine learning to approximate quantum dynamics equations via Wick rotation \cite{wick1954properties}. By defining imaginary time as $\tau = it$, the quantum dynamics equations shed their inherent fluctuation characteristics, thereby transforming into classical diffusion reaction equations under the regime of imaginary time. This transformation enables a more tractable approximation on classical computational platforms.
		\begin{equation}
			\frac{\partial \psi(x, \tau)}{\partial \tau} = \left( D \frac{\partial^2}{\partial x^2} - \zeta(x) \right) \psi(x, \tau)
		\end{equation}
		Following Wick's rotation, the zeroth-order Taylor approximation of the generalized objective function, denoted by $\zeta \left( x \right)$, undergoes a transformation from the quantized free electron equation to the classical diffusion equation:
		\begin{equation}
			\frac{\partial \psi(x, \tau)}{\partial \tau} = D \frac{\partial^2 \psi(x, \tau)}{\partial x^2}
		\end{equation}
		The dynamics of the algorithm degenerates from wave packet dispersion to a classical diffusion process (e.g., Figure 2). The Green's function for the one-dimensional diffusion equation is Gaussian function :
		\begin{equation}
			\psi(x, \tau) = \frac{1}{(4\pi D \tau)^{3/2}} e^{-\frac{x^2}{4D \tau}}
		\end{equation}
		The physical interpretation of the Green's function is to describe the diffusion process of a point source. This is defined as the probability that a point located at the origin 0 will diffuse to a position $x$ at the $\tau$ moment. The behaviour of sampling in the solution space of an optimisation problem can be described using the diffusion of a point source. The classical diffusion process can be implemented in functional optimisation problems using population Gaussian random walk sampling. The parameter D is here the diffusion coefficient. A larger value of D corresponds to a larger kinetic energy of the system and a larger standard deviation of the corresponding Gaussian sampling.
		\begin{figure}[H]
			\centering
			\includegraphics[width=1\linewidth]{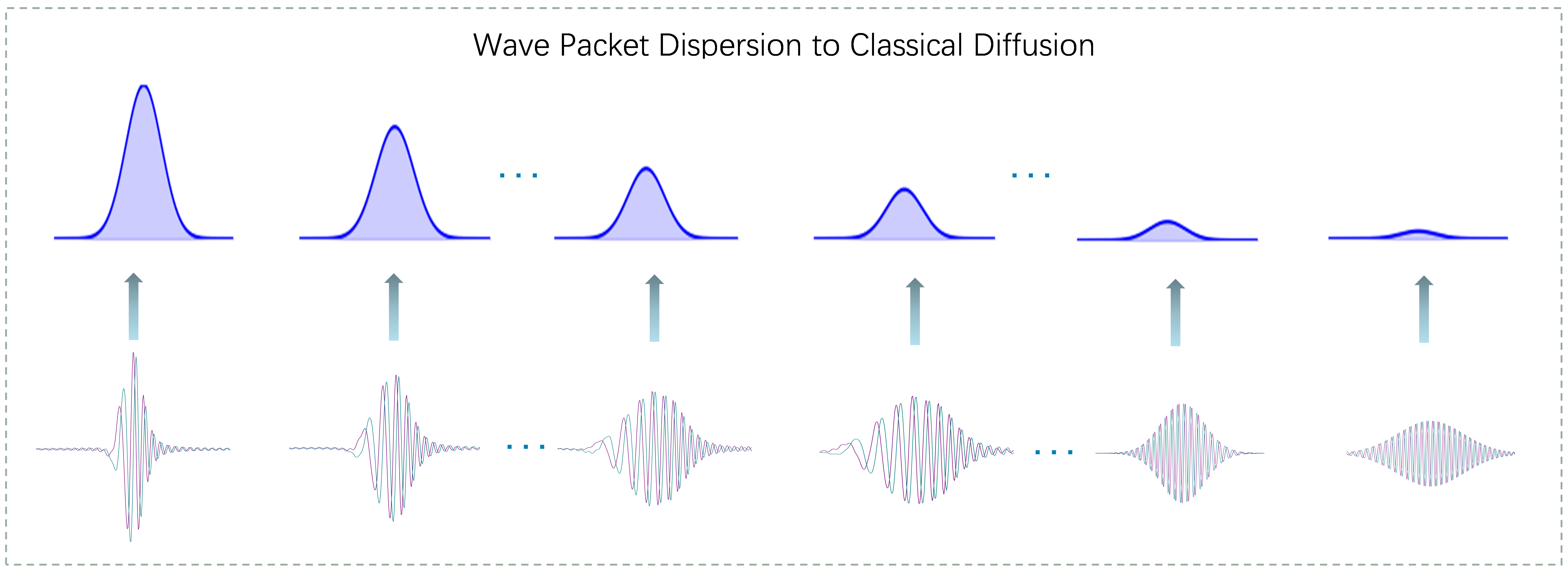}
			\caption{Transformation From Wave Packet Dispersion to Classical Diffusion}
			\label{fig:3}
		\end{figure}
		It can be demonstrated that the fundamental dynamical process of machine learning is a Gaussian stochastic diffusion process in the parameter space when the information about the generalised objective function, denoted by $\zeta \left( x \right)$, is completely unknown. 
		The first-order Taylor approximation of the generalised objective function, $\zeta \left( x \right)$, after Wick's rotation is the Fokker-Planck equation.
		\begin{equation}
			\frac{\partial \psi(x, \tau)}{\partial \tau} = \left( D \frac{\partial^2}{\partial x^2} - \frac{\partial \zeta(x)}{\partial x} \right) \psi(x, \tau)
		\end{equation}
		The drag term $\frac{\partial \zeta \left( x \right)}{\partial x}$ denotes the process of acquiring the derivative of the generalized objective function $\zeta \left( x \right)$ at the current sampling location during machine learning iterations. Unlike the diffusion equation, which addresses the scenario where the generalized objective function $\zeta \left( x \right)$ is entirely unknown, the Fokker-Planck equation describes the dynamics when the derivative at the current sampling position is accessible, allowing for gradient descent. Given that the derivative of $\zeta \left( x \right)$ in parameter space cannot be directly defined, it can be approximated and estimated through double sampling. In machine learning, the Fokker-Planck equation delineates the gradient descent process based on Gaussian sampling diffusion. Each learning iteration for a sample during machine learning can be mapped to a single gradient descent operation in the parameter space for the generalized objective function $\zeta \left( x \right)$.
		\section{Application of Quantum Dynamics Theory in Machine Learning}
		\subsection{Convergence Analysis of Machine Learning}
		The establishment of quantum dynamical equations for machine learning provides reliable theoretical support for the study of the iterative process of machine learning. This can be used to analyse and explain some core operations and algorithms in machine learning. the quantum dynamics equation is employed to analyse the convergence of the iterative process of machine learning.\\
		The iterative process of machine learning can be described as a time-evolutionary process. The general solution of the machine learning quantum dynamics equation can be expressed in the following form:
		\begin{equation}
			\psi(x,t) = \sum_{n} c_{n} \phi_{n}(x) \exp(-i E_{n} t)
		\end{equation}
		The general solution formula indicates that the solution to the quantum dynamical equations is a superposition of states formed by the probabilistic superposition of a series of states at different energy levels. Upon the measurement of this superposition of states, the superposition will probabilistically collapse to one of the states. The energy $E_n$ in the generic solution corresponds to the kinetic energy $E_d$ plus the potential energy $E_p$ of the system:
		\begin{equation}
			E_{n} = E_{d} + E_{p}
		\end{equation}
		The general solution of the classical diffusion reaction equation after the quantum kinetic equation undergoes Wick's rotation is as follows:
		\begin{equation}
			\psi(x, \tau) = \sum_{n} c_{n} \phi_{n}(x) \exp(-E_{n} \tau)
		\end{equation}
		In order to solve the optimisation problem of finding the minimum, it is necessary to obtain the base energy $E_0$.\\
		The exponential component of the generalised solution following Wick's rotation results in the states corresponding to each energy level undergoing exponential decay over the course of the evolution of $\tau$. The larger the value of the energy $E_n$, the faster the decay. Consequently, the terms corresponding to the ground state energy $E_0$ will be preserved with a higher probability, i.e., the superior solution will be preserved with a high probability. As $\tau$ approaches infinity, the following is obtained:
		\begin{equation}
			\begin{aligned}
				& \lim_{\tau \to \infty} \psi(x, \tau) \\
				& = \lim_{\tau \to \infty} \sum_{n} c_{n} \phi_{n}(x) \exp(-E_{n} \tau) \\
				& \approx c_{0} \phi_{0}(x) \exp(-E_{0} \tau)
			\end{aligned}
		\end{equation}
		The state of lowest energy is given by $c_{0}\phi_{0}(x)\exp(-E_{0}\tau)$, where $c_{0}$ is a constant and $\phi_{0}$ is the wave function.\\
		The results demonstrate that the iterative process of machine learning converges to the ground state, i.e., the global optimal solution, as $\tau$ approaches infinity. Furthermore, the Wick rotation of the quantum dynamics equations for machine learning is shown to be an effective method for expressing the evolutionary convergence properties of the algorithm to the ground state, despite the loss of volatility. The time evolution process of thermodynamics acts as a low-pass filter for the energy of the system, with low-energy states being retained with a high probability.\\
		The convergence of this time evolution and the physical significance of the coefficients $D$ related to kinetic energy in the kinetic equations leads to the conclusion that the kinetic process of machine learning should consist of the following two basic iterative processes:\\
		Firstly, an annealing process is employed which continuously decreases the value of the coefficient $D$ in order to gradually reduce the kinetic energy of the system. In practice, this process corresponds to a gradual reduction of the sampling step size in order to obtain a lower ground state energy $E_0$. This implies that the machine learning algorithm should have some kind of multiscale learning process. Secondly, there is a process of temporal evolution towards the ground state of the system with the same kinetic energy. This is an iterative evolution process that can be observed in the gradient descent operation using the Softmax or Sigmoid criterion.
		\subsection{Derivation of Softmax and Sigmoid}
		Similarly, the widely used Softmax and Sigmoid functions in machine learning can be derived from the generalised formulae of the classical diffusion reaction equations.\\
		It is assumed that the general solution of the classical diffusion reaction equation after the quantum dynamics equation undergoes a Wick rotation consists of states at $n$ energy levels:
		\begin{equation}
			\psi(x, \tau) = \sum_{i=1}^{n} c_{i} \phi_{i}(x) \exp(-E_{i} \tau)
		\end{equation}
		The state corresponding to the energy level $E_k$ at time $\tau$ is given by $c_k\varphi_k(x)\mathrm{exp}(-E_k\tau)$.\\
		The initial moment of algorithm evolution, designated as $\tau = 0$, represents the state of the system at that point in time.
		\begin{equation}
			\psi(x, 0) = \sum_{i=1}^{n} c_{i} \phi_{i}(x)
		\end{equation}
		Given that time $\tau$ is in the exponential position, the size of each term is primarily determined by the exponential term as it evolves over time. Conversely, states with higher energy will decay more rapidly as they evolve over time. As the algorithm evolves over time, the exponential term will become dominant, rendering the effect of non-exponential terms insignificant. Consequently, the probability of the appearance of a state with energy $E_k$ can be approximated as:
		\begin{equation}
			P_{E_{k}} \approx \frac{\exp(-E_{k} \tau)}{\sum_{i=1}^{n} \exp(-E_{i} \tau)}
		\end{equation}
		This equation represents the Softmax function with respect to time. The Softmax function will evolve dynamically with time $\tau$, with the probability of a state with low energy increasing if $\tau$ gradually increases. As $\tau$ approaches infinity, the system will converge to the lowest energy state, $E_0$. Here, Softmax implements a low-pass filtering of the system's energy, thereby preserving the less energetic states with greater probability.\\
		A two-energy approximation to the system is made if the system energy is reduced to two by simplifying the system into one cloth. These energies are designated by the symbols $E_a$ and $E_b$, respectively. Assuming that $E_a < E_b$, $E_a$ is the low-energy state corresponding to the superior solution and $E_b$ is the high-energy state corresponding to the inferior solution, the probability of occurrence of the low-energy state $E_a$ is given by:
		\begin{equation}
			\begin{aligned}
				P_{E_{a}} & \approx \frac{\exp(-E_{a} \tau)}{\exp(-E_{a} \tau) + \exp(-E_{b} \tau)} \\
				& = \frac{1}{1 + \exp[-(E_{b} - E_{a}) \tau]} \\
				& = \frac{1}{1 + \exp(-\Delta E \tau)}
			\end{aligned}
		\end{equation}
		The probability function resulting from the two-energy approximation is a time-dependent sigmoid function, whose shape undergoes a transformation over time. If the difference in energy \(\Delta E\) is constant, the probability of selecting the low-energy state \(P_{E_{a}}\) increases with the passage of time \(\tau\). Furthermore, the probability of selecting the low-energy state \(P_{E_{a}}\) tends to one as \(\tau\) approaches infinity.
		\subsection{Quantum Dynamics Interpretation of the Diffusion Model}
		The diffusion model is a machine learning algorithm based on thermodynamic theory, which was proposed in 2015. It has demonstrated excellent performance and potential applications in a number of fields, including generative artificial intelligence. The diffusion model employs a Gaussian diffusion method, whereby Gaussian noise is added to the image in a stepwise manner, and then the noisy image is denoised in a similar manner in order to learn the network parameters.\\
		The relationship between quantum dynamics and diffusion processes was initially explored by Anderson in 1975, who proposed the DMC method for solving the ground state wave function of a molecule using a random walk. This method was developed based on the isomorphism between the Schrödinger equation and the diffusion equation \cite{anderson1975random}. The DMC is considered a standard method for obtaining the ground state energy and wave function of a quantum system. Furthermore, the DMC method can be used to compute the ground state wave function of a quantum system if an optimisation problem is transformed into a quantum problem \cite{kosztin1996introduction}. The ground state wave function of an optimisation problem can be computed using the DMC method if the problem is transformed into a quantum problem. DMC exploits the isomorphism (isomorph) between the Schrödinger equation and the diffusion equation by simulating the diffusion process to drive the wave function to evolve towards the ground state step by step. This process is used to achieve the search for the globally optimal solution of the objective function \cite{haghighi2017full}.\\
		The machine learning quantum dynamics equations were analysed, and it was found that the 0th order Taylor approximation of the generalised objective function after the classical approximation requires the machine learning algorithm to sample the generalised objective function $\zeta \left( x \right)$ in a Gaussian manner in the parameter space. Furthermore, the 1st order Taylor approximation requires the machine learning algorithm to sample the generalised objective function $\zeta \left ( x \right)$ twice in order to obtain an estimate of the gradient. However, the generalized objective function, represented by the symbol $\zeta(x)$, is only a formal description of the objective function in the parameter space. Consequently, it is not possible to accurately define its sampling neighbourhood in the same way as is possible for a function optimization problem. This is because, in the case of a generalized objective function, direct Gaussian sampling in the parameter space is not permitted. Consequently, the Gaussian noise diffusion and denoising process of the diffusion model in the learning space can be regarded as an approximation of the Gaussian sampling process in the parameter space in the learning space.
		As illustrated in Figure 4, the diffusion model indirectly implements the Gaussian sampling diffusion process of the machine learning algorithm in the parameter space through the Gaussian noise diffusion process in the learning space, thereby resolving the issue that the machine algorithm is unable to perform Gaussian sampling in the parameter space directly. Consequently, the iterative process of the diffusion model is consistent with the theoretical framework of quantum dynamics in machine learning. In accordance with the theory of quantum dynamics, the diffusion model can be regarded as the 0th and 1st order iterative operations of the generalised objective function under the classical approximation of machine learning quantum dynamics. In the context of the diffusion model, each denoising iteration is analogous to a Gaussian sampling operation within the generalised parameter space. Furthermore, a sequence of one additive and denoising iteration for a sample can be viewed as a sequence of Gaussian diffusion samples for a sample point within the parameter space.
		\begin{figure}[H]
			\centering
			\includegraphics[width=0.95\linewidth,height=0.72\textheight]{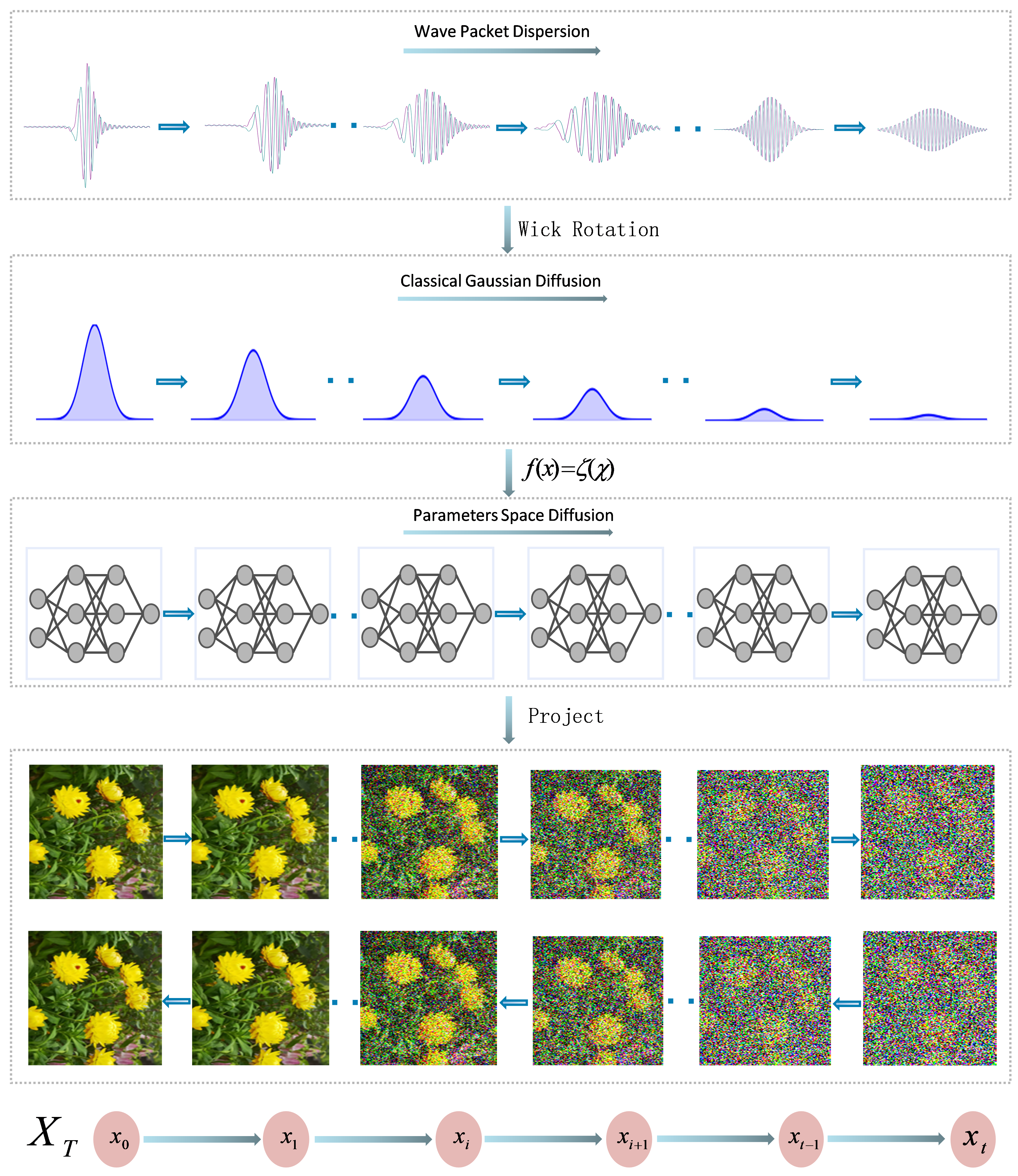}
			\caption{Quantum Dynamics Interpretation of The Diffusion Model}
			\label{fig:4}
		\end{figure}
		As illustrated in Fig. 5, the diffusion model indirectly implements the Gaussian sampling diffusion process in the parameter space by eliminating the learning process of Gaussian noise, thereby completing the mapping of the learning space to Gaussian sampling and gradient acquisition in the parameter space. The Gaussian sampling in the parameter space represents a classical approximation of the wave packet dispersion process in quantum kinetic theory.
		\begin{figure}[H]
			\centering
			\includegraphics[width=0.9\linewidth,height=0.2\textheight]{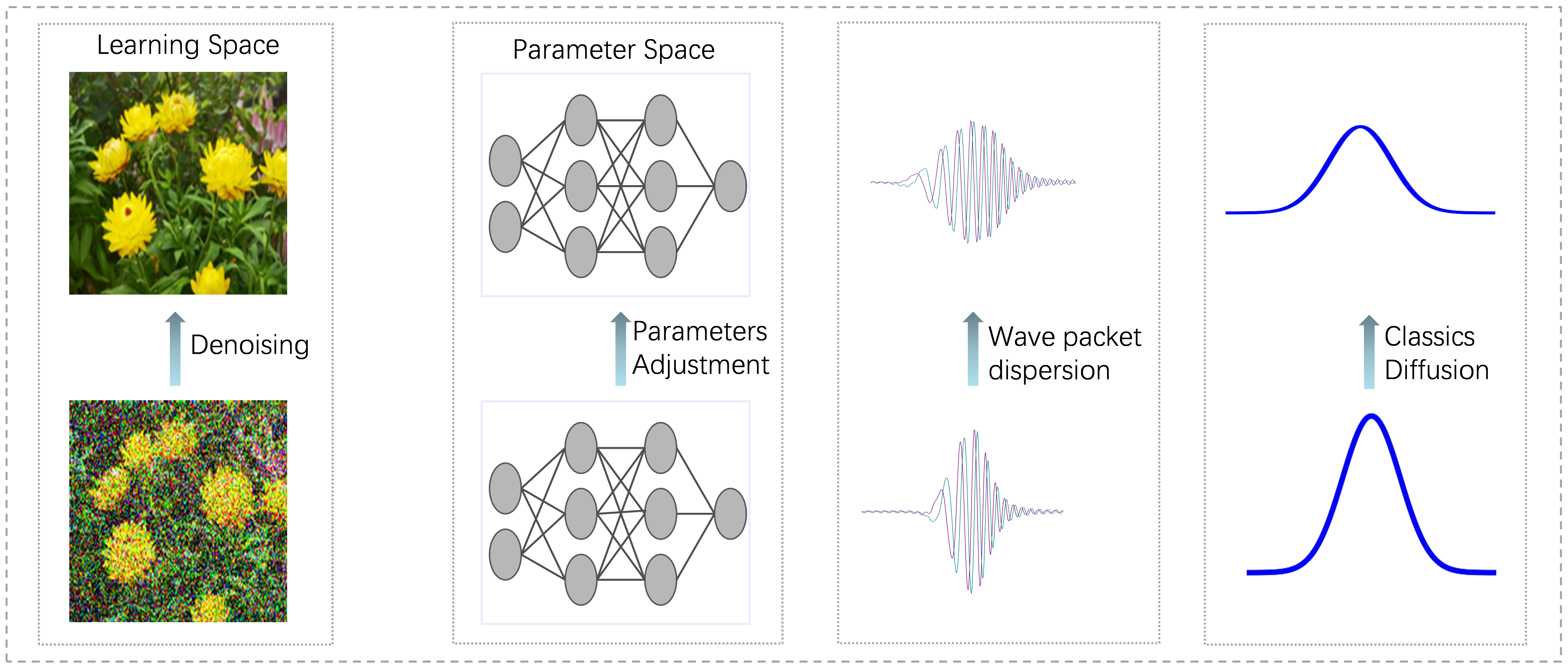}
			\caption{Sampling Mapping of Parameter Space}
			\label{fig:5}
		\end{figure}
		Concurrently, the quantum dynamics of machine learning indicate that the iterative process of machine learning should also have an annealing process that gradually reduces the kinetic energy. This suggests that it may be necessary to add noises with different standard deviations from large to small in the diffusion model to achieve the annealing of the system. The efficacy of the approach of incorporating multi-scale noise in the context of diffusion models is currently being demonstrated experimentally \cite{jeong2023multiscale}. Consequently, the complete diffusion model must also comprise two iterative looping processes. The first of these is the annealing process, which involves a reduction in the standard deviation of Gaussian noise. The second is the diffusion evolution process.
		\section{Conclusion}
		This paper establishes the quantum dynamics equation for the iterative process of machine learning based on Schrödinger's equation and relates it to the thermodynamics equation through Wick rotation.
		The quantum dynamics equation for machine learning was used to obtain the basic iterative process of machine learning through Wick rotation and Taylor approximation. The generalised solution of the equation was then used to obtain the Softmax and Sigmoid functions commonly used in machine learning, thereby providing a physical and statistical explanation of the significance of these functions. Finally, the quantum kinetic equations were used to give a kinetic interpretation of the basic iterative process of the diffusion model.
		The establishment of quantum dynamics equations for machine learning transforms the iterative process of machine learning into time-containing partial differential equations, which achieves an accurate mathematical description of the machine learning process. This enables the use of mature theoretical systems in quantum mechanics and mathematics to carry out research on machine learning. The work presented in this paper offers a novel research perspective and a theoretical foundation for the precise establishment of machine learning. Furthermore, it is anticipated that this theoretical framework will provide a foundation for the implementation of artificial intelligence algorithms on quantum computers in the future. computers in the future.
		
		\bibliographystyle{unsrt}  
		\bibliography{references}

\begin{thebibliography}{10}

\bibitem{metropolis1953equation}
N.~Metropolis, A.~Rosenbluth, M.~Rosenbluth, A.~H. Teller, and E.~Teller.
\newblock Equation of state calculations by fast computing machines.
\newblock {\em The Journal of Chemical Physics}, 21(6):1087--1092, 1953.

\bibitem{kirkpatrick1983optimization}
S.~Kirkpatrick, C.~D. Gelatt, and M.~P. Vecchi.
\newblock Optimization by simulated annealing.
\newblock {\em Science}, 220(4598):671--680, 1983.

\bibitem{wang2013multiscale}
Peng Wang, Yan Huang, Chao Ren, and Youming Guo.
\newblock Multi-scale quantum harmonic oscillator for high-dimensional function global optimization algorithm.
\newblock {\em Acta Electronica Sinica}, 41(12):2468--2473, 2013.

\bibitem{wang2022review}
Peng Wang and Fang Wang.
\newblock A review of intelligent optimization algorithms in quantum perspective.
\newblock {\em Journal of University of Electronic Science and Technology (Natural Science Edition)}, 51(1):2--15, 2022.

\bibitem{wang2023quantum}
Peng Wang and Gang Xin.
\newblock Quantum theory of intelligent optimization algorithms.
\newblock {\em Acta Automatica Sinica}, 49(11):2396--2408, 2023.

\bibitem{wang2024convergence}
Fang Wang and Peng Wang.
\newblock Convergence of the quantum dynamics framework for optimization algorithm.
\newblock {\em Quantum Information Processing}, 23(3):66, 2024.

\bibitem{johnson2011quantum}
M.~W. Johnson, M.~H.~S. Amin, and S.~Gildert.
\newblock Quantum annealing with manufactured spins.
\newblock {\em Nature}, 473(7346):194--198, 2011.

\bibitem{sohl2015deep}
Jascha Sohl-Dickstein, Eric Weiss, Niru Maheswaranathan, and Surya Ganguli.
\newblock Deep unsupervised learning using nonequilibrium thermodynamics.
\newblock In {\em Proceedings of the 32nd International Conference on Machine Learning}, pages 2256--2265. PMLR, 2015.

\bibitem{song2020score}
Yang Song, Jascha Sohl-Dickstein, Diederik~P. Kingma, Abhishek Kumar, Stefano Ermon, and Ben Poole.
\newblock Score-based generative modeling through stochastic differential equations.
\newblock {\em arXiv preprint arXiv:2011.13456}, 2020.

\bibitem{xin2021multiscale}
Gang Xin, Peng Wang, and Yuwei Jiao.
\newblock Multiscale quantum harmonic oscillator optimization algorithm with multiple quantum perturbations for numerical optimization.
\newblock {\em Expert Systems With Applications}, 185:115615, 2021.

\bibitem{jin2021multiscale}
Jin Jin and Peng Wang.
\newblock Multiscale quantum harmonic oscillator algorithm with guiding information for single objective optimization.
\newblock {\em Swarm and Evolutionary Computation}, 65:100916, 2021.

\bibitem{wick1954properties}
G.~C. Wick.
\newblock Properties of bethe-salpeter wave functions.
\newblock {\em Physical Review}, 96(4):1124--1134, 1954.

\bibitem{anderson1975random}
J.~B. Anderson.
\newblock A random walk simulation of the schr{\"o}dinger equation: H+3.
\newblock {\em The Journal of Chemical Physics}, 63(4):1499--1503, 1975.

\bibitem{kosztin1996introduction}
I.~Kosztin, B.~Faber, and K.~Schulten.
\newblock Introduction to the diffusion monte carlo method.
\newblock {\em American Journal of Physics}, 64(5):633--644, 1996.

\bibitem{haghighi2017full}
M.~K. Haghighi and A.~Lüchow.
\newblock Full wave function optimization with quantum monte carlo and its effect on the dissociation energy of fes.
\newblock {\em The Journal of Physical Chemistry A}, 121(32):6165--6171, 2017.

\bibitem{jeong2023multiscale}
Jongheon Jeong and Jinwoo Shin.
\newblock Multi-scale diffusion denoised smoothing.
\newblock In {\em Thirty-seventh Conference on Neural Information Processing Systems}, 2023.

\end{thebibliography}

	\end{document}